\documentclass{elsart5p}

\usepackage{natbib}
\usepackage{graphicx}
\usepackage{amssymb}

\begin{document}

\begin{frontmatter}

\title{The Venus ground-based image Active Archive: a database of amateur observations of Venus in ultraviolet and infrared light}


\author{Geert Barentsen\corauthref{gb}},
\ead{Geert@Barentsen.be}
\corauth[gb]{Corresponding author. Tel.: +31 71 565 4828}
\author{Detlef Koschny}
\ead{Detlef.Koschny@esa.int}

\address{Research and Scientific Support Department (RSSD),
ESA/ESTEC, SCI-SM, Keplerlaan 1, 2201 AZ Noordwijk, The Netherlands}

\begin{abstract}
The Venus ground-based image Active Archive is an online database designed to collect ground-based images of Venus in such a way that they are optimally useful for science. The Archive was built to support ESA's Venus Amateur Observing Project, which utilises the capabilities of advanced amateur astronomers to collect filtered images of Venus in ultraviolet, visible and near-infrared light. These images complement the observations of the Venus Express spacecraft, which cannot continuously monitor the northern hemisphere of the planet due to its elliptical orbit with apocentre above the south pole. We present the first set of observations available in the Archive and assess the useability of the dataset for scientific purposes.
\end{abstract}
\end{frontmatter}

\section{Introduction}
In recent years, amateur planetary observers have become increasingly capable to obtain datasets that are useful for scientific analysis. Very few amateurs, however, actually contribute to science because they are not actively encouraged or supported by the scientific community to do so. This is unfortunate because the continuous observing capabilities of amateur astronomers can be of value to study certain time-variable planetary phenomena. In particular, worldwide amateur observations can be used to augment spacecraft observations that are confined to narrow fields of view or brief observing time slots. For example, amateur images of Jupiter from the {\em Planetary Virtual Observatory \& Laboratory} (PVOL\footnote{PVOL website: http://www.pvol.ehu.es/}) have recently been used to complement observations of the Hubble Space Telescope for a publication in Nature~\citep{San2008}, thereby demonstrating the potential of amateur-professional collaborations in planetary sciences.

In this paper, we present the {\em Venus Amateur Observing Project} (VAOP) which was initiated to encourage and support amateur observers of Venus to contribute to the ongoing study of the planet. In particular, we discuss the {\em Venus ground-based image Active Archive} (VAA) software which was developed to collect the observing results in a proper scientific way. We present the initial dataset that was collected in this archive and assess the useability of the dataset for scientific analysis.

\section{The Venus Amateur Observing Project}
The {\em Venus Amateur Observing Project} (VAOP\footnote{VAOP website: http://sci.esa.int/VAOP}) was first launched in March 2006, shortly before the arrival of the Venus Express spacecraft in its orbit around Venus. The project focuses on utilising the capabilities of advanced amateur astronomers to obtain narrowband filtered images of the atmosphere of Venus in ultraviolet, visible and near-infrared light. Amateurs have been able to obtain such images in recent years thanks to the advent of low light level CCD cameras and cheap bandpass filters.

\begin{figure*}
\begin{center}
\includegraphics*[width=\linewidth]{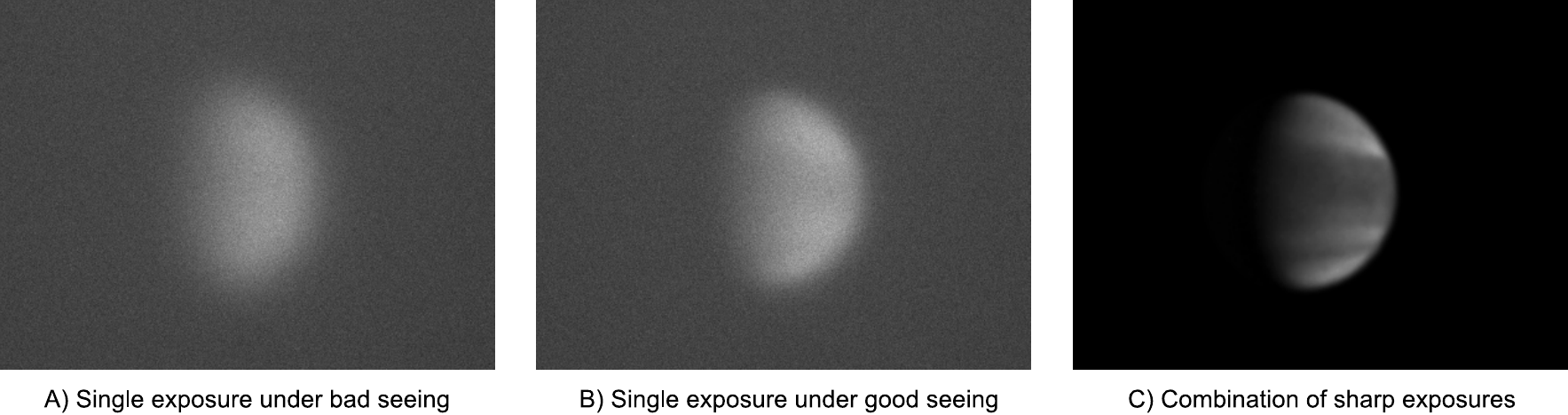}
\end{center}
\caption{Illustration of the Lucky Imaging technique. Image A (left) shows an unsharp exposure of 33~ms taken at a moment of heavy atmospheric turbulence. Image B (middle) shows a sharp exposure of 33~ms taken at a moment with little atmospheric turbulence. Image C (right) shows the result after selecting, aligning and adding 7980 exposures together to increase the signal-to-noise ratio. The images were taken on 15 April 2007 at 15:32 UT using a DMK 21AF04 video camera equipped with a Baader UV bandpass-filter (centered around 350~nm). The telescope used was a 12.5~inch f/4.8 Newtonian equipped with a $5\times$ Televue Powermate barlow lens. The Giotto software was used to process the final image. The images were taken by amateur astronomer Ralf Gerstheimer, who is a member of a German Venus observing team~\citep{Kow2007}.}
\label{fig:fig1}
\end{figure*}

Amateur astronomers obtain their images using an increasingly popular technique from video-astronomy called {\em Lucky Imaging}. In this technique, the blurring effect caused by atmospheric turbulence (commonly known as {\em seeing}) is circumvented by obtaining a large set of very short exposures (100~ms or less) and selecting the sharpest frames that were least affected by the atmospheric blurring. These sharp exposures are then aligned and added together to increase the signal-to-noise ratio. Using this technique, it is possible to ``freeze'' the atmospheric turbulence and reach the diffraction limit of the telescope. These processing steps are typically performed using freely available software such as Registax\footnote{Registax website: http://www.astronomie.be/registax} and Giotto\footnote{Giotto website: http://www.videoastronomy.org}. An illustration of the technique is given in Figure~\ref{fig:fig1}. The technique is described in more detail in~\cite{Law2006}.

The scientific objective of the Amateur Observing Project is to complement the observations of the {\em Venus Express} (VEX) spacecraft. The VEX instrument package includes the {\em Venus Monitoring Camera} (VMC) and the {\em Visible and Infrared Thermal Imaging Spectrometer} (VIRTIS), which both observe the planet in ultraviolet, visible and near-infrared light~\citep{VMC,VIRTIS}. The spacecraft has a high inclination elliptical orbit around Venus with an apocentre of 66~000~km above the south pole and a pericentre of 250~km near the north pole and a period of 24 hours~\citep{VEX}. This orbit implies that the spacecraft moves slowly around the south pole, where it obtains extended nadir observations of the entire southern hemisphere, and moves very fast over the equatorial and northern latitudes, where it obtains high-resolution close-up snapshot views. Ground-based observations are therefore useful to complement the VEX observations with images of the equatorial and northern regions of the planet, which cannot be continuously monitored by the spacecraft. In addition, it is important to compare ground-based observations with simultaneous spacecraft observations, because this allows us to extend our understanding of ground-based observations made before and after the VEX mission. We will provide a comparison of an amateur image with VEX data in Section~\ref{firstresults}.

In parallel to the Amateur Observing Project, about twenty teams of professional astronomers joined efforts to perform targeted observations of Venus in the {\em Coordinated Ground-based Venus Observation Campaign} between 23 May 2007 and 9 June 2007\footnote{Coordinated campaign website: http://sci.esa.int/venusexpress}. The Amateur Observing Project is complementary to this campaign, because most professional teams do not produce the same kind of continuous full-disk image series that amateur observers do.

\section{The Active Archive database}
Amateur astronomers commonly submit their observations of Venus to organizations such as the {\em Association of Lunar and Planetary Observers} (ALPO) and the {\em British Astronomical Association} (BAA). These observations are typically archived as JPEG-compressed files with metadata information added as text to the image. For scientific data analysis, however, these JPEG images with embedded text are not very useful for several reasons. First, JPEG is a lossy compression method that introduces artefacts into the image. Second, the metadata information is often incomplete and cannot be queried or processed in an automated way because it is embedded in the pixels. Third, observers tend to combine multiple images in one file as a mosaic or by producing a processed color image, whereas scientists may want to access the raw data.

For these reasons, it was proposed to create a dedicated {\em Venus ground-based image Active Archive} (VAA\footnote{Active Archive website: http://www.rssd.esa.int/vaa}) to support the Amateur Observing Project. The VAA is a web-application designed to collect images of Venus in such a way that they are optimally useful for science. The Archive collects uncompressed files in the TIFF- or FITS-format and asks the observer to complete a set of predefined metadata keywords. This is done in an interactive way: the metadata keywords are automatically validated to avoid input errors and to ensure that all the necessary information is really there.

When the images are submitted, they undergo a preliminary quality review by an administrator. The images that are accepted by the administrator are then stored in a database that can be queried online and allows images to be downloaded in TIFF or FITS-format (even if the original image was not a FITS file, the Archive will automatically produce a FITS-file with the provided metadata keywords embedded in the header to facilitate automated analysis routines). The images are publically accessible, but the Archive specifies usage conditions to ensure that proper credit is given to the observers. Another feature of the database is that it also stores the light transmission curves for the bandpass filters used by the observers, which is a common demand from scientists.

The Active Archive was initially developed to support the Amateur Observing Project, but it was implemented in a generic way to be suited for professional data as well. Moreover, it is possible to ingest the data from the Archive into ESA's official planetary data archive, the {\em Planetary Science Archive} (PSA), although this would require a proper scientific review of the observations.

\section{First set of images available in the Active Archive}
\label{firstresults}
The ground-based image Active Archive was officially launched in November 2007. By the end of 2007, a total of 128 high-quality full-disk images of Venus were submitted to the archive, 14 in the infrared wavelength band, 114 in the ultraviolet. The telescopes used ranged from 8~inch Newtonians, Schmidt- or Maksutov-Cassegrains; up to the 80~cm reflector of the Public Observatory Munich. Mostly, video cameras with 8 bit to 12~bit dynamic range were used. Different commercially available UV filters were employed, e.g., the Baader UV filter which is a bandpass filter centered around 350~nm and having a width of 100~nm. Most images were taken with effective focal lengths of 6 to 10~m, resulting in an image scale of 0.2 to 0.5" per pixel. Depending on the distance to Venus, this results in an image scale of typically 60 to 200~km per pixel.
\begin{figure}
\begin{center}
\includegraphics*[width=\linewidth]{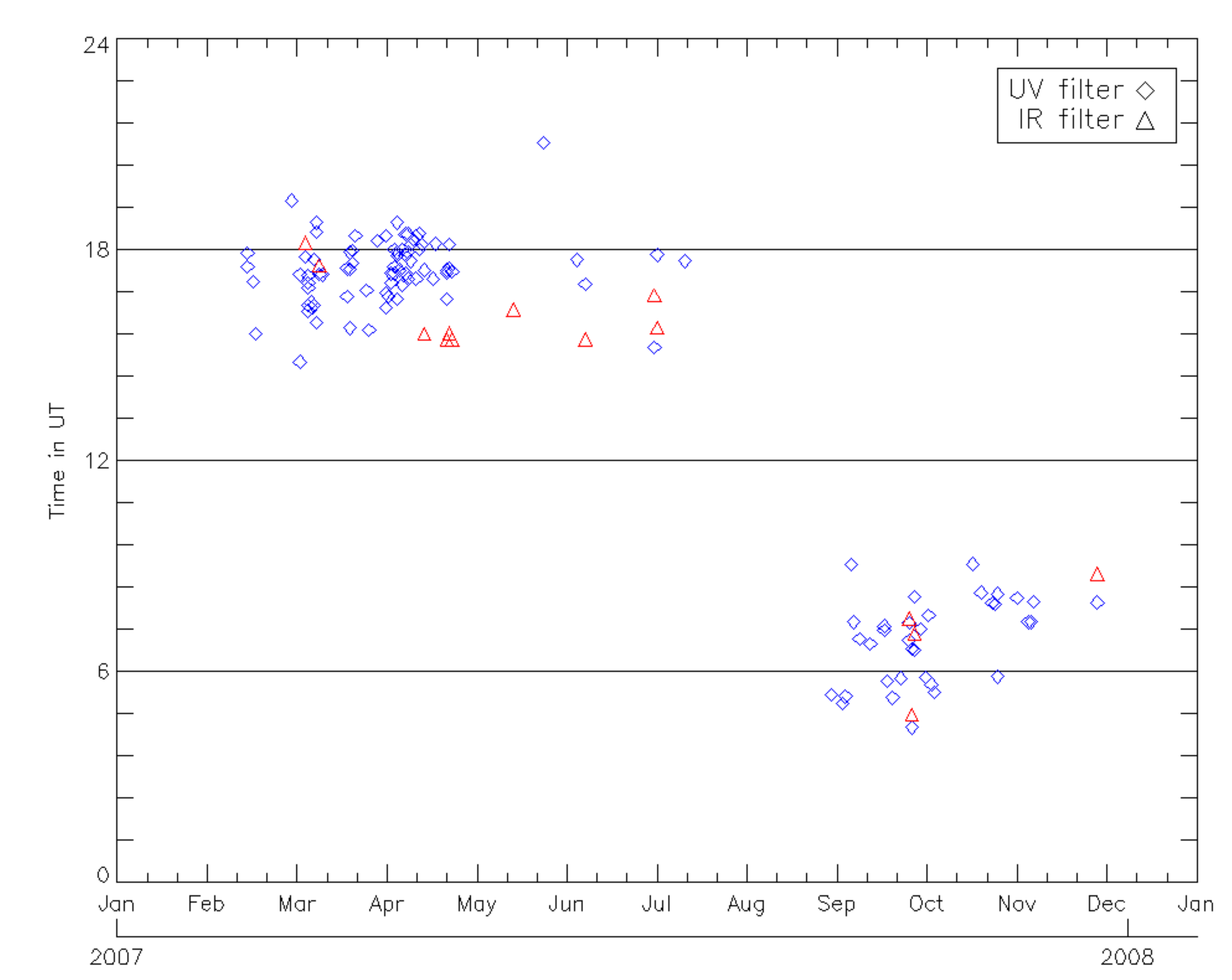}
\end{center}
\caption{Time of all the images taken versus the date, as submitted to the Venus ground-based image Active Archive by the end of 2007. The morning (top left) and evening (bottom right) elongations of Venus are obvious. }
\label{fig:fig2}
\end{figure}

Figure~\ref{fig:fig2} shows the time of all images taken versus date in the year 2007. The morning and evening visibilities near the eastern and western elongations are obvious. The best temporal coverage is currently available in the period March/April 2007. Note that observations in the visible light have not been submitted in 2007, this wavelength band is not popular with amateurs because it shows little or no features.

\begin{figure}
\begin{center}
\includegraphics*[width=\linewidth]{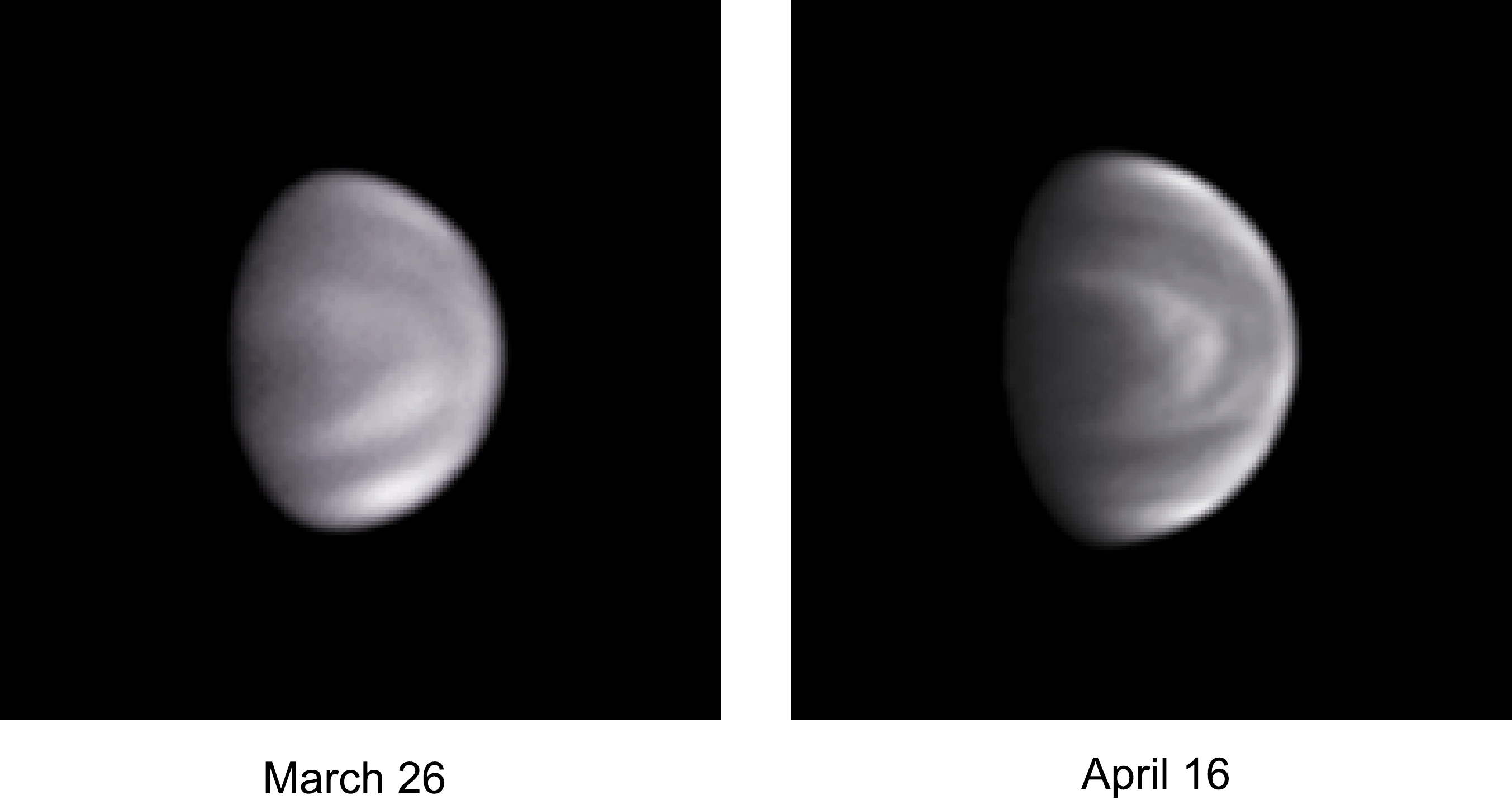}
\end{center}
\caption{Example images by amateur astronomers Gabriele and J\"org Ackermann from Germany. The images were taken on 26 March 2007 at 15:46 UT (left) and 16 April 2007 at 17:10 UT (right) using a 180~mm Maksutov-Cassegrain telescope equipped with a DMK 21AF04 video camera and a Baader UV filter. A total of 9500 sharp frames, each taken with 66~ms exposure time, were combined with the Registax software to obtain the final result. Features in the atmosphere of Venus can clearly be seen. These images and their metadata keywords are available in the Active Archive.}
\label{fig:fig3}
\end{figure}

\begin{figure}
\begin{center}
\includegraphics*[width=\linewidth]{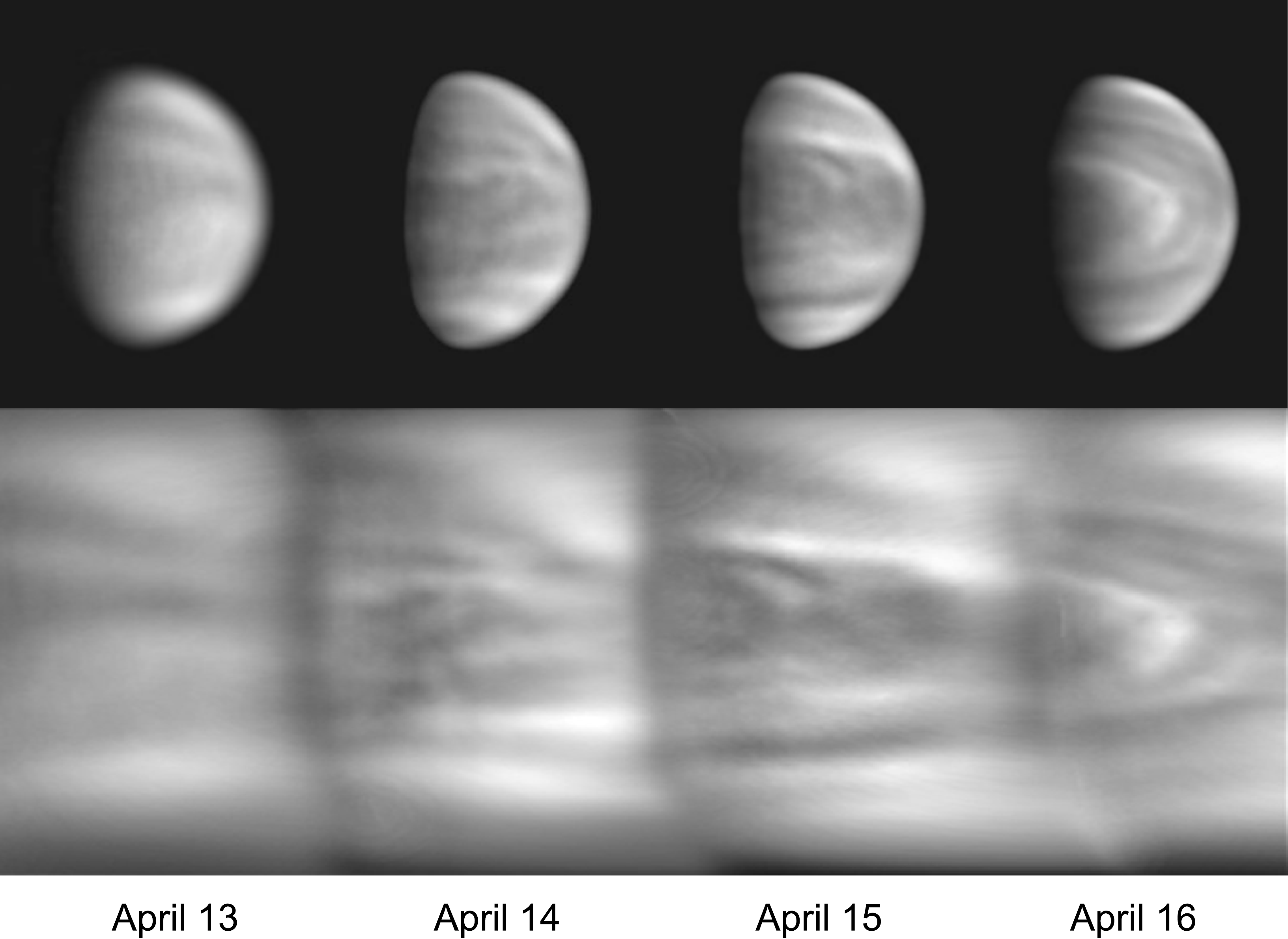}
\end{center}
\caption{Example images by amateur astronomer Bernd G\"ahrken~\citep{Kow2007}. The images were taken from 13 to 16 April 2007, each near 18:00 UT, using an 80~cm reflector telescope equipped with a Philips ToUCam 740 webcam with a modified chip and a Schuler UV filter. On average, 4500 sharp frames were combined to obtain each image. The images were then manually reprojected and assembled together by the observer to obtain a cloud map (bottom), which is possible because the cloud tops of Venus circle the planet about every four to five earth days. These images and their metadata keywords are available in the Active Archive.}
\label{fig:fig4}
\end{figure}

\begin{figure}
\begin{center}
\includegraphics*[width=\linewidth]{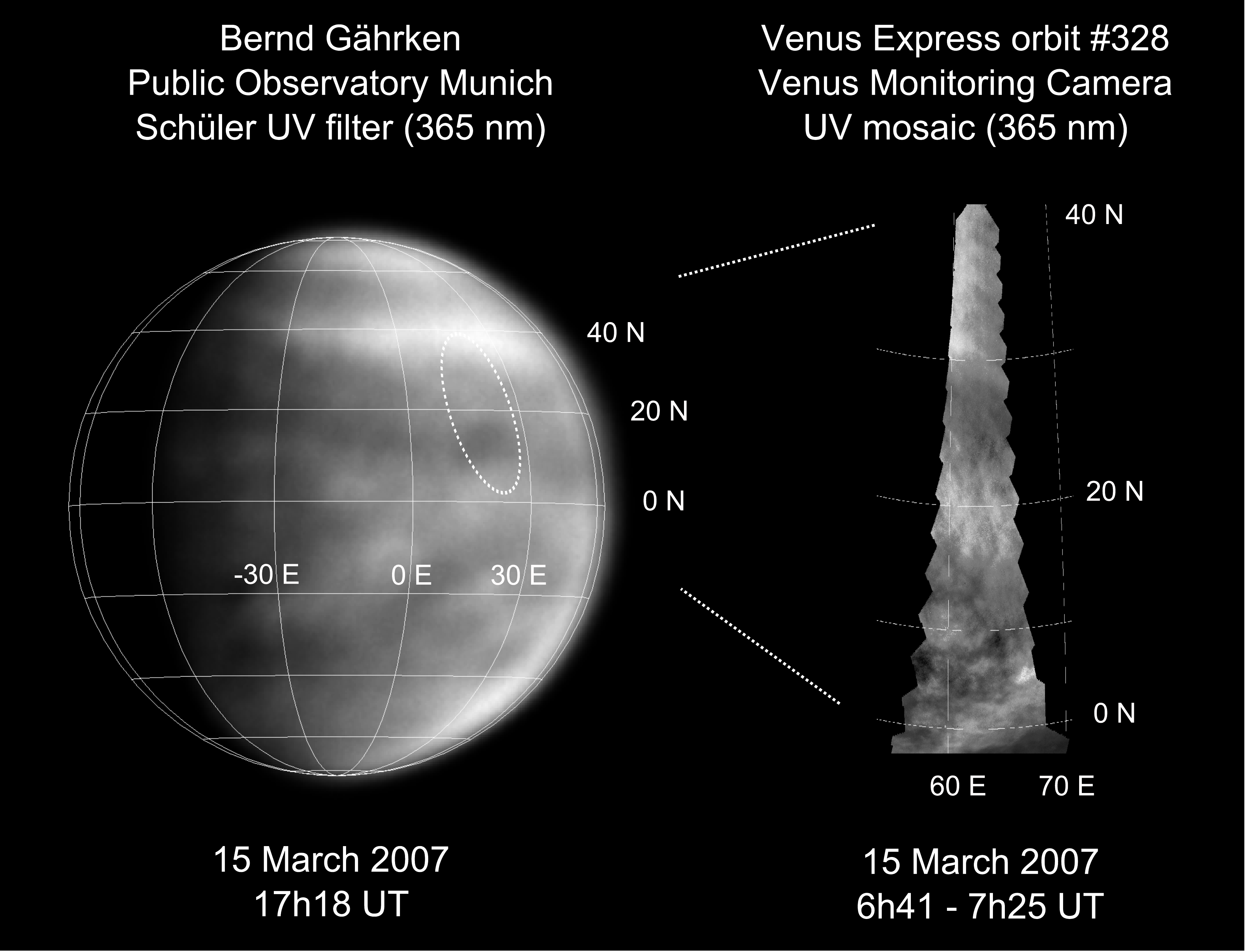}
\end{center}
\caption{Comparison of an amateur image (left) with a mosaic of dayside images taken by the Venus Express VMC camera (right). It is obvious that the low-resolution full-disk amateur image can provide a context for the high-resolution close-up images obtained by the spacecraft when it flies low over the northern hemisphere. The VMC images were taken 10.5 hours earlier than the amateur image. In this period, the ultraviolet cloud top features seen on the VMC image would have moved approximately 35 degrees west if we assume the clouds at these latitude regions to rotate the planet westward in a period of 4.5 days~\citep{VMCNature}. Indeed, on the amateur image we see a pattern near 25 degrees east longitude (dotted line) that resembles the cloud structure seen earlier that day by the VMC camera near 60 degrees east longitude.} 
\label{fig:fig5}
\end{figure}

Examples of high-quality ultraviolet images taken with one of the smaller telescopes, a 180~mm Maksutov-Cassegrain, are shown in Figure~\ref{fig:fig3}. Example of ultraviolet images taken with the biggest telescope, the 80~cm reflector of the Public Observatory Munich, are shown in Figure~\ref{fig:fig4}. We took one of the images from this telescope and compared it with an ultraviolet mosaic from the VMC camera on Venus Express, taken on the same day. The result of this comparison can be seen in Figure~\ref{fig:fig5}.


\section{Assessment of the useability of the dataset for science}
To assess the useability of the dataset for scientific purposes, we used the image data from March and April 2007 to study the brightness of the Venus' polar clouds with respect to the complete disk of Venus versus time. This time period shows the most complete coverage in the data set, typically there is at least one image per day available. Only images in the UV wavelength range were used. The images were taken with different UV filters: Schuler UV, Baader U, or special combinations of Baader U and a Schott BG40 filter to increases the contrast of the images. All filters are bandpass filters centered around 350~nm and with a width of 100 nm. Individual differences between the filters were considered insignificant and all images are treated the same.

Most images were marked as being dark- and flatfield corrected. Some images were not - for those an artificial background image was produced. We took the mean value of the dark sky background in the four corners of the image, then interpolated linearly for the complete image. This image was subtracted from the raw data. No flat fielding was applied to these images. For each image, three regions of interest were defined: an ellipse inscribed Venus' disk; an ellipse or a polygon around the area close to the north pole; another ellipse or polygon around the area close to the south pole. Then, the ratio between the mean pixel value of the pole areas and the complete Venus disk was determined after subtracting the background. 

\begin{figure}
\begin{center}
\includegraphics*[width=\linewidth]{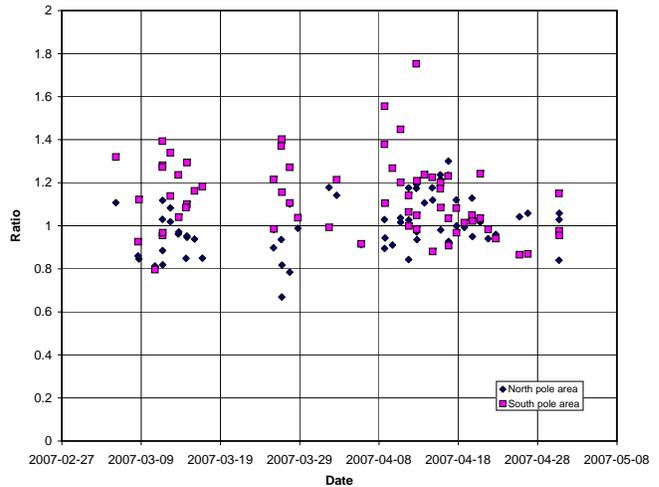}
\end{center}
\caption{Ratio of the brightness of the north (diamonds) and south (squares) pole areas to the complete disk of Venus in March and April 2007. In March the south pole area appears to have been brighter on average, whereas the north pole area was not.}
\label{fig:fig6}
\end{figure}

The result is shown in Figure~\ref{fig:fig6}. The ratio of the brightness of the north pole area to the complete Venus disk is shown with diamonds, the same for the south pole is shown with squares. The data seems to indicate that in March the south pole area has, in almost all cases, a ratio larger than one, meaning it was brighter than the average Venus. The north pole area, on the other hand, is typically around 1, i.e. on average not brighter or darker than the Venus disk. Also between different observers using different recording setups, the ratios show the same trend. For the month of April, there is no indication of a trend visible.\\
When analysing the data in more detail, it became clear that one observer who contributes a significant amount of data in April, but not in March, does not take sufficient care in defining the keyword values required by the data archive. In particular, it is not in all cases clear where the north direction in the image is, towards the top or the bottom of the image. Thus, some ratio values may be the wrong way around.

In addition, no absolute calibration of the images is available and the stretch which was applied to the data is not known. Thus, the absolute values of the atmospheric brightness between different observers cannot be compared. For the same observer, the mean brightness of Venus stays roughly the same, however, between different observers this is not the case. Thus, care has to be taken when absolute values are used.

Also, handling the calibration of the images seems not to be clear to all observers. In some images there is indication that an artifical background value was subtracted, or, in some cases, it looks like the background was simply covered with an artifical value. In these cases the administrator needs to provide feedback to the observers.

The lack of data in certain time periods can mainly be attributed to bad weather. Thus, it would be important to augment the data set with observers from other locations (currently, most of the observers are from Germany or the Benelux countries). 

In summary, we conclude the following:
\begin{itemize}
	\item There is possible evidence that the south pole area of Venus was brighter than average in March 2007.
	\item More work needs to be done to inform amateurs about proper calibration techniques and make them aware of the importance to correctly document the necessary meta information like the image orientation. This may be done by providing well-documented observing procedures and keyword template forms.
\end{itemize}
Thus, in particular if a proper data validity checking mechanism is in place and additional documentation is created, data from the Venus Active Archive can contribute to Venus science.

\section{Discussion and further work}
The Active Archive demonstrates that amateur observers are willing to make an extra effort to submit their observations in a way that the data is optimally useful for science. However, we experienced that it is necessary to have an administrator available to actively monitor incoming data and to provide feedback to observers. We would like to stress that this is a time-consuming task that should not be underestimated. Therefore, to let this project run beyond the lifetime of the Venus Express mission, it would be necessary to obtain the permanent support of a research institute or amateur organization. 

In further work, the Archive could be extended to support observations of other planets and comets. For example, continuous monitoring of a comet's tail by amateurs could yield interesting time series on the interaction of the tail with the solar wind. In general, we suggest future professional observing campaigns and planetary spacecraft missions to consider including amateur observing projects to augment their data. The Active Archive was implemented in a generic way to support such projects. At the time of writing this article, however, there was no plan or funding to actually do so. 

Finally, we note that it may be interesting to integrate the Archive with similar survey projects such as the archive of outer planet images in the {\em Planetary Virtual Observatory \& Laboratory} (PVOL) of the {\em International Outer Planets Watch} (IOPW\footnote{IOPW website: http://www.ehu.es/iopw}). Their effort doest not enforce strict rules on the availability of metadata keywords and the use of lossless data compression, but the goals are similar.

\section{Conclusion}
The Venus ground-based image Active Archive is an online database collecting amateur observations of Venus in such a way that they are useful for science. The Archive is succesfully being used to complement the observations of the VMC and VIRTIS instruments of the Venus Express spacecraft with full-disk observations in ultraviolet and infrared light. We assessed the useability of the dataset for scientific purposes and find that the data may contribute to Venus science if a proper data validity checking mechanism is in place. The Archive was initially aimed at amateur astronomers, but can also be used for data from professional observers. Moreover, data ingestion from the Archive into the ESA Planetary Science Archive (PSA) can be done but would require a proper scientific review. 

\section*{Acknowledgements}
The authors wish to thank the following people for providing ground-based Venus observations to the Venus ground-based image Active Archive in 2007: Gabriele Ackermann, J\"org Ackermann, Bart Declercq, Bernd G\"ahrken, Willem Kivits, Silvia Kowollik, Jean-Pierre Prost, Arnaud van Kranenburg and Mario Weigand. We also wish to thank Ralf Gerstheimer for providing the Lucky Imaging illustration. Special thanks go to Silvia Kowollik for providing valuable feedback during the development of the Archive.

We greatly appreciate the support of the VMC team, in particular Wojciech Markiewicz and Richard Moissl from MPS in Katlenburg-Lindau and the VMC Co-I team at DLR Berlin. They provided the VMC mosaic and supported our comparison with the ground-based images.

\end{document}